\documentclass[12pt,a4paper]{panl}
\usepackage{cite}
\usepackage{wrapfig}
\usepackage{graphicx}
\usepackage{amssymb}
\usepackage{amsfonts}
\usepackage{amsmath}
\usepackage{longtable}
\usepackage{rotating}
\usepackage{lscape}
\usepackage{epsfig}
\usepackage{multirow}

\originalTeX
\begin{document}

\title{ Double spin correlations in the reaction dd→ pnpn and in the pn- elastic scattering
}
\maketitle
\authors{Yu.\,Uzikov$^{a,b,c}$\footnote{E-mail: uzikov@jinr.ru},
A.\,Temerbayev$^{b,}$\footnote{E-mail: adastra.77@email.ru}}
\setcounter{footnote}{0}
\from{$^{a}$\, V.P.Dzhelepov  Laboratory of Nuclear Problems, Joint Institute for Nuclear Researches, Dubna, Russia }
\from{$^{b}$\, M.V. Lomonosov Moscow State University, Physics  Faculty, Moscow, Russia }
\from{$^{c}$\, Dubna  State University, Dubna, Russia }
\from{$^{d}$\,L.N. Gumilyov Eurasian  National University, Astana, Republic of Kazakhstan}

\begin{abstract}

\vspace{0.2cm}

An unexpectedly large double spin correlation with a sharp dependence on energy,
found in elastic  proton-proton scattering at large angles in the c.m.s. energy range  $\sqrt{s_{pp}}= 3 \div  5.5$ GeV,
may be associated with the formation of exotic octoquark resonances
at the thresholds of the strangeness and charm production.
For a deeper understanding of the dynamics of this process
it is important to measure
double spin asymmetry in elastic $pn$ scattering
in the same kinematics. A possibility of obtaining such  data from the reaction $d d\to pnp$ with two polarized deuteron beams
on NICA  SPD is considered in this paper.
\end{abstract}
\vspace*{6pt}



\label{sec:intro}
\section*{Introduction}
\label{Uz-ANN}

Elastic proton- nucleon  scattering  at the c.m.s. angle $\theta_{cm}=90^\circ$  and   the invariant mass of the $pN$ system of $\sqrt{s}= 3 - 5$ GeV is very  interesting because
the  transferred 4-momentum $|t|$ in this case  is   rather large
and corresponds to very small relative distances between
nucleons, $r_{pN}\sim \hbar/\sqrt{|t|}\leq 0.1$fm. This is the region  of spatial overlap of nucleons, and therefore  it is necessary to use quark degrees of freedom in describing the dynamics of this process. Indeed, the measured differential cross section of elastic $pp$ scattering at the fixed angle $\theta_{cm}\sim 90^\circ$ at these and higher energies, on average, follows a power dependence on $s$ in  the form  of $d\sigma^{pp}/dt({s},\theta_{cm})\sim s^{-10}$ \cite{Allaby:1968zz,Akerlof:1967zz,Perl:1969pg,Stone:1977jh}, which is consistent with the constituent  quark counting rules (QCR), that  is a prediction of both the hypothesis of self-similarity \cite{Matveev:1973ra} and perturbative quantum chromodynamics (pQCD) \cite{Brodsky:1973kr}. However, there is also a deviation from this prediction, that are small but clearly visible oscillations  observed in the energy dependence of the differential cross section in the energy range under discussion $\sqrt{s}=3\div 6$ GeV
\cite{Allaby:1968zz,Akerlof:1967zz,Perl:1969pg,Stone:1977jh}.

For inclusive processes at  hard  momentum transfer, quantum chromodynamics predicts the phenomenon  of color transparency (CT), which consists in the fact that
hadrons participating in a hard  process in its initial or final state, when interacting with a nuclear medium, experience a weakening of absorption in the medium compared to a conventional soft process \cite{Mueller,Brodsky}.
The reason for this weakening of the interaction is that hadrons can effectively participate in a hard  process only being in compact, point-like configurations, which provide a high momentum transfer,
and at the same time the color dipole moments of these hadrons, being proportional to the transverse dimensions of the hadrons $\sqrt{<r^2_\bot>}$,  decrease in size. For this reason, the cross section of the interaction of the hadron with the nucleus also decreases in proportion to  its size $\sqrt{<r^2_\bot>}$. The color transparency was established experimentally in the processes with the production  of mesons. For processes with baryons, experimental indications  to  this phenomenon in hadron interactions are quite elusive. In particular, in the reaction of proton knock-out  from  nuclei by protons A(p,2p)B, the effect of attenuation of interaction with the nuclear medium takes place
at certain initial energies, but with a further increase of energy, the absorption of protons in the nucleus increases again, contrary to the prediction of the  CT model \cite{Mardor:1998zf,Aclander:2004zm}.

Experimental data on elastic $pp$ scattering of transversely polarized protons  at the angle
$\theta_{cm}=90^\circ$ and  laboratory momenum  $p_{lab} =1.2 \div 11.8$ GeV/c demonstrate a clear dependence of the cross section on the choice of parallel or antiparallel mutual orientation of
polarization vectors of colliding protons \cite{Crabb:1978km,Bhatia:1982sy,Lin:1978xu,Crosbie:1980tb,Court:1986dh}.
The measured  double spin asymmetry $A_{NN}$ of  this process strongly depends on energy and this is in  contradiction with the prediction of the perturbative QCD, which provides  an energy-independent value $A_{NN} =1/3$.
In   Ref. \cite{Brodsky:1987xw}, an explanation of this contradiction between theory and experiment for $A_{NN}$ was proposed based on the assumption of the excitation of the octoquark resonances $uuuds\bar s uud$ and $uuudc\bar c uud$ at the   strangeness $(s\bar s)$  and charm
$(c\bar c)$ thresholds, respectively. At the same time, the interference between the perturbative QCD  amplitude and the non-perturbative resonant amplitude allowed the authors of  Ref. \cite{Brodsky:1987xw} to describe not only the experimental data on $A_{NN}$, but also the above-mentioned oscillations in the energy dependence of the differential cross section of the $pp$ scattering at the angle $\theta_{cm}\sim 90^\circ$, as well
as the noted above unexpected behavior of color transparency  in the reaction $A(p,2p)B$. However, there is another explanation for these oscillations \cite{Ralston:1982pa} and the behavior of the CT  in the reaction $(p,2p)$ -- the mechanism of the nuclear filter \cite{Ralston:1988rb}. Furthermore, recent  measurements of the reaction$^{12}$C(e,ep)X
at transferred momenta squared  $Q^2=8-14\, (GeV/c)^2$
\cite{Bhetuwal:2020jes} showed  not CT effect. This raises questions about the analysis carried out in papers
\cite{Brodsky:1987xw,Ralston:1988rb} (see details in the works of \cite{Larionov:2022gvn} and references therein).

In this regard, it is of great interest to obtain an independent information from the measurement of the spin correlation $A_{NN}$ in elastic $pn$ scattering at  the angle
$\theta_{cm}\sim 90^\circ$ and the same energies as for the  $pp$ scattering. The point  is that the spin-isospin structure of the $pn\to pn$ process differs significantly from the $pp\to pp$ process by the presence of an isoscalar channel in $pn$ scattering \cite{Rekalo:2002ck}. The  available data on the $A_{NN}$ in elastic $pn$ scattering in the energy region  under discussion are obtained  only at  two points along the transferred momentum at the  initial momentum  of 6 GeV/c \cite{Crabb:1979nh}.
In this paper, we consider the possibility of obtaining new information about the double spin asymmetry $A_{NN}$ in an elastic
$pn$-scattering from the reaction $d d\to pnpn$, in which both  deuteron beams will be  polarized in planned experiment at  NICA SPD.

\section{Elements of formalism}

\subsection{Transition amplitude of the reaction $dd\to pnpn$.}
\label{Matrfi}
\begin{figure}[t]
\begin{center}
\vspace{-3mm}
\includegraphics[width=70mm]{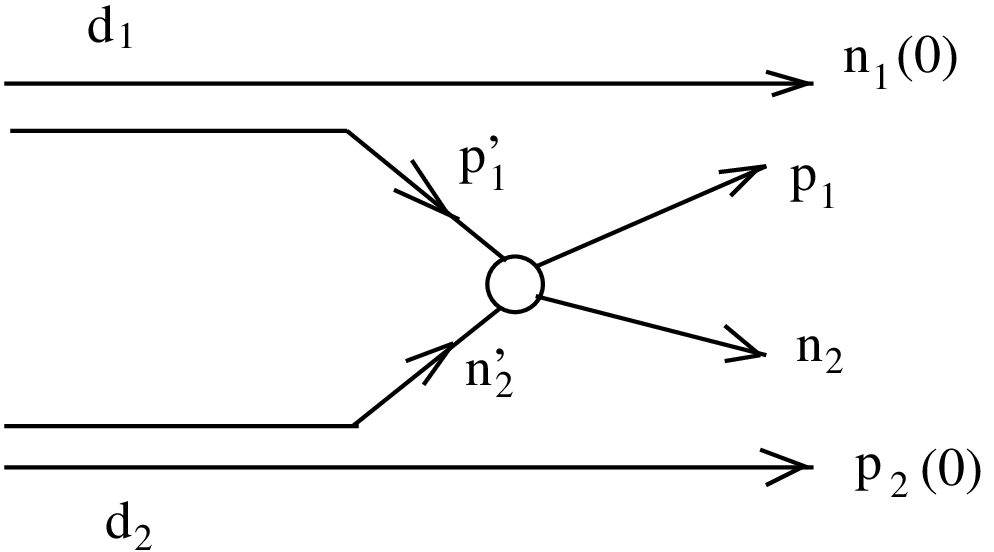}
\vspace{-3mm}
\caption{The  mechanism of the reaction $dd\to pnpn$.}
\end{center}
\labelf{fig1}
\vspace{-5mm}
\end{figure}
\vspace{0.5cm}
Assuming the double pole mechanism of the reaction $dd\to pnpn$ (Fig. \ref{fig1}),
the transition  amplitude can be written as
\begin{eqnarray}
 M_{fi}=\sum _{\sigma_{n_2'} \sigma_{p_1'}}
 \frac {M(d_1\to n_1 p_1'){ i}
 T_{NN}(p_1' n_2'\to p_1n_2 ){ i} M(d_2\to n_2'p_2)}
 {(p_{n_2'}^2-m^2+i\epsilon)(p_{p_1'}^2-m^2+i\epsilon)},
 \label{Mfi}
\end{eqnarray}
where
$T_{NN}(p_1' n_2'\to p_1n_2)$
is the t-matrix of the elastic NN-scattering;
$m$ is the
 mass of the  nucleon;
the product of the nucleon propagator $(p_{p_1'}^2-m^2+i\epsilon)^{-1}$ and
the amplitude of the virtual decay of the deuteron $d_1\to n_1 p_1'$  in Eq. (\ref{Mfi})
 can be written as
\begin{eqnarray}
\frac{ M(d_1\to n_1p_1')}{({p_{p_1'}}^2-m^2+i\epsilon)}
=
-<\chi_{\sigma_{n_1}}\chi_{\sigma_{p_1'}}|
\phi_{\lambda_1}({\bf q})>
{{2m}}.
\end{eqnarray}
Here we have the overlap of the deuteron wave function in momentum space for the spin state with
the spin projection $\lambda_1$,
$\phi_{\lambda_1}({\bf q})$, with the Pauli spinors  of the nucleons
$\chi_{\sigma_{n_1}}$
and $ \chi_{\sigma_{p_1'}}$,
where $\sigma_i$ is
 the {\i}-th nucleon spin projection. Then
 the transition matrix element  for the reaction  $dd\to pnpn$ takes the form
\begin{eqnarray}
 M_{\lambda_1\lambda_2}^{\sigma_{n_1}\sigma_{p_1}\sigma_{n_2}\sigma_{p_2}}
=
\sum _{\sigma_{n_2'} \sigma_{p_1'}}
<\chi_{\sigma_{p_1}}(p_1) \chi_{\sigma_{n_2}}(n_2)
|t_{pn}|\chi_{\sigma_{p_1'}}(p_1) \chi_{\sigma_{n_2'}}(n_2)>\nonumber \\
\times
<\chi_{\sigma_{n_1}}\chi_{\sigma_{p_1'}}|\phi_{\lambda_1}({\bf q})>
<\chi_{\sigma_{n_2'}}\chi_{\sigma_{p_2}}|\phi_{\lambda_2}({\bf q})>.
\end{eqnarray}
 Here  we take into account  only   the S-wave component of the deuteron
 wave function $u(q)$ that gives the following relations:
\begin{eqnarray}
<\chi_{\sigma_{n_1}}\chi_{\sigma_{p_1'}}|\phi_{\lambda_1}({\bf q})>
=u(q_1)
(\frac{1}{2}\sigma_{n_1}\frac{1}{2}\sigma_{p_1'}|1\lambda_1), \nonumber \\
 <\chi_{\sigma_{p_2}}\chi_{\sigma_{n_2'}}|\phi_{\lambda_2}({\bf q})>
 =
 u(q_2)
 (\frac{1}{2}\sigma_{p_2}\frac{1}{2}\sigma_{n_2'}1|1\lambda_2),
\end{eqnarray}
where standard notations for the Clebsh-Gordan coefficients are used.

The  differential cross section of the reaction for definite spin projections of initial deutreons $\lambda_1$  and $\lambda_2$  can be written as
\begin{eqnarray}
d\sigma_{\lambda_1\lambda_2}=\frac{1}{9}K\sum_{{\sigma_{n_1}\sigma_{p_1}\sigma_{n_2}\sigma_{p_2}}}
| M_{\lambda_1\lambda_2}^{\sigma_{n_1}\sigma_{p_1}\sigma_{n_2}\sigma_{p_2}}|^2
\nonumber \\
=\frac{1}{9}K\sum_{{\sigma_{n_1}\sigma_{p_1}\sigma_{n_2}\sigma_{p_2}}}
|<\chi_{\sigma_{p_1}}(p_1) \chi_{\sigma_{n_2}}(n_2)
|t_{NN}|\chi_{\sigma_{p_1'}=\lambda_{1}-\sigma_{n_1}}(p_1)
\chi_{\sigma_{n_2'}=\lambda_2-\sigma_{p_2}}(n_2)>
\nonumber \\
\times
(\frac{1}{2}\sigma_{n_1}\frac{1}{2}\sigma_{p_1'}
|1\lambda_1)^2
(\frac{1}{2}\sigma_{p_2}\frac{1}{2}
\sigma_{n_2'}
|1\lambda_2)^2u(q_1)^2u(q_2)^2,
\label{dsiglamlam}
\end{eqnarray}
where $K$ is a kinematic factor.

\subsection{Double spin correlation in the reaction $d d\to pnpn$.}

\vspace{0.5cm}
Unpolarized  deuteron beam has different components of the z-projections of the
spin states $\lambda =\pm 1, 0$ in equal portions:
$N_{+}=N_{-}=N_{0}$. The spin  ``up'' ($N_{\uparrow}$) and spin ``down'' ($N_{\downarrow}$) beams are
prepared by spin flip of the states $\lambda=1$ and $\lambda=-1$, respectively:
\begin{eqnarray}
N_{\uparrow}=N_{+} + N_{+} +N_0, \nonumber \\
N_{\downarrow}=N_{-}+N_{-}+N_0.
\label{Nupdown}
\end{eqnarray}
The polarization of the beam  along the axis $OY$ which is directed  along the magnetic field, is given as the following asymmetry
\begin{eqnarray}
P_{Y}=\frac{N_{\uparrow}-N_{\downarrow}}{N_{\uparrow}+N_{\downarrow}}=\frac{N_{+} - N_{-}}{ N_{+} + N_{0}+N_{-}}
\label{Py}
\end{eqnarray}
The tensor polarization (alignment) is
\begin{eqnarray}
P_{YY}=\frac{{ N}_{\lambda =+1}+{ N}_{\lambda =-1}-2{ N}_{\lambda =0}}
{{ N}_{\lambda =+1}+{ N}_{\lambda =-1}+{N}_{\lambda =0}}.
\label{Pyy}
\end{eqnarray}
One can see from
Eqs. (\ref{Nupdown}),(\ref{Py}) and  \ref{Pyy})
that for
$N_{\uparrow}$ ($N_{\downarrow}$) the vector polarization of the beam is
$P_y=+\frac{2}{3}$ ($P_y=-\frac{2}{3}$),
 whereas the tensor polarization is zero: $P_{YY}=0$.

Let us consider
collision of two deuteron beams in two cases with different beam polarizations
$P_y(1)$ and $P_y(2)$:
1) $  \,P_y(1)=+\frac{2}{3}, P_y(2)=+\frac{2}{3}$
and
2) $  \,P_y(1)=+\frac{2}{3}, P_y(2)=-\frac{2}{3}$.
The event count for the  first case is denoted as ${\cal N}_{\uparrow\uparrow}$ and for the second as ${\cal N}_{\uparrow\downarrow}$.
Then the  double spin asymmetry is  defined as
\begin{eqnarray}
 A_{YY}^{dd}=\frac{{\cal N}_{\uparrow\uparrow}-{\cal N}_{\uparrow\downarrow}}
 {{\cal N}_{\uparrow\uparrow}+{\cal N}_{\uparrow\downarrow}}.
 \label{Ayy}
\end{eqnarray}
In terms of $d\sigma_{{\lambda_1}{\lambda_2}}$ one has for the event counts
\begin{eqnarray}
{\cal N}_{\uparrow\uparrow}= L(2\cdot 2d\sigma_{++}+2d\sigma_{+0}+ 2d\sigma_{0+}+d\sigma_{00}),\nonumber \\
{\cal N}_{\uparrow\downarrow}=L(2\cdot 2d\sigma_{+-}+2d\sigma_{+0}+ 2d\sigma_{0-}+d\sigma_{00}),
\label{Nupup}
\end{eqnarray}
where the factor $L$ accounts for  luminosity and detector efficiency.
Using Eq. (\ref{dsiglamlam}), one can find
\begin{eqnarray}
d\sigma_{\lambda_1=+1\lambda_2=+1}=f\sum_{\sigma_{p_1}\sigma_{n_2}}
|<\chi_{\sigma_{p_1}}(p_1) \chi_{\sigma_{n_2}}(n_2)
|t_{NN}|\chi_{\sigma_{p_1'}=+\frac{1}{2}}(p_1)
\chi_{\sigma_{n_2'}=+\frac{1}{2}}(n_2)>|^2, \nonumber \\
d\sigma_{\lambda_1=+1\lambda_2=-1}=
f\sum_{\sigma_{p_1}\sigma_{n_2}}
 |<\chi_{\sigma_{p_1}}(p_1) \chi_{\sigma_{n_2}}(n_2)
|t_{NN}|
\chi_{\sigma_{p_1'}=+\frac{1}{2}}(p_1)
\chi_{\sigma_{n_2'}=-\frac{1}{2}}(n_2)>|^2,
\label{dsig}
\end{eqnarray}
where $f=u(q_1)^2u(q_2)^2$. We use  below the relation
\begin{eqnarray}
 \sum_{\sigma_{p_1}\sigma_{n_2}}
 |<\chi_{\sigma_{p_1}}(p_1) \chi_{\sigma_{n_2}}(n_2)
|t_{NN}|\chi_{\sigma_p}(p) \chi_{\sigma_n}(n)>|^2=\nonumber \\
=
\sum_{\sigma_{p_1}\sigma_{n_2}}
 | <\chi_{\sigma_{p_1}}(p_1) \chi_{\sigma_{n_2}}(n_2)
|t_{pn}|\chi_{-\sigma_p}(p) \chi_{-\sigma_n}(n)>|^2.
\label{R3}
\end{eqnarray}
 Eq. (\ref{R3}) can be obtained
making rotation on the angle $\theta=\pi$ over the axis orthogonal  to the quantization axis, and using  rotational invariance. Taking into account Eq.(\ref{R3}), we find from Eqs. (\ref{dsiglamlam}) the following relations:
\begin{eqnarray}
d\sigma_{0,1}=d\sigma_{0,-1}= d\sigma_{-1,0}=
d\sigma_{1,0}=d\sigma_{0,0}.
\label{dsigmll}
\end{eqnarray}
As a result,  the double spin-correlation  given by Eq. (\ref{Ayy})  can be present in the form
\begin{eqnarray}
 A_{YY}^{dd}=\frac{d\sigma_{+1,+1}-d\sigma_{+1,-1}}{d\sigma_{+1,+1}-d\sigma_{+1,-1} +\frac{5}{2}d\sigma_{0,0}}=
 \frac{4}{9}A_{YY}^{NN},
\end{eqnarray}
 where $A_{YY}^{NN}$ is the double spin-correlation in elastic $NN$-scattering, that is the subprocess  of the reaction $dd\to pnpn$ (Fig. \ref{fig1}):
\begin{eqnarray}
 A_{YY}^{NN}=
\frac {F_{\chi_{+\frac{1}{2}}(p) \chi_{+\frac{1}{2}}(n)} -F_{\chi_{+\frac{1}{2}}(p) \chi_{-\frac{1}{2}}(n)}}{F_{\chi_{+\frac{1}{2}}(p) \chi_{+\frac{1}{2}}(n)}+ F_{\chi_{+\frac{1}{2}}(p) \chi_{-\frac{1}{2}}(n)}};
\end{eqnarray}
here
$F_{\chi_{+\frac{1}{2}}(p) \chi_{+\frac{1}{2}}(n)}= \newline
=\sum_{\sigma_{p_1}\sigma_{n_2}}
|<\chi_{\sigma_{p_1}}(p_1) \chi_{\sigma_{n_2}}(n_2)
|t_{NN}|\chi_{\sigma_{p_1'}=+\frac{1}{2}}(p_1)
\chi_{\sigma_{n_2'}=+\frac{1}{2}}(n_2)>|^2$\newline
and
$F_{\chi_{+\frac{1}{2}}(p) \chi_{-\frac{1}{2}}(n)}=\newline
=\sum_{\sigma_{p_1}\sigma_{n_2}}
|<\chi_{\sigma_{p_1}}(p_1) \chi_{\sigma_{n_2}}(n_2)
|t_{NN}|\chi_{\sigma_{p_1'}=+\frac{1}{2}}(p_1)
\chi_{\sigma_{n_2'}=-\frac{1}{2}}(n_2)>|^2$.

\subsection{Spin-correlation parameter $C_{y,y}$. }
 Using  standard notations \cite{Ohlsen:1972zz}, the differential  cross section  $I$ for  interaction
  of two spin-1 particles can be written as
   \begin{eqnarray}
    I=I_0(1+\frac{3}{2}P_yA_y+\frac{3}{2}P_y^TA_y^T+\frac{9}{4}P_yP_y^TC_{y,y}),
    \label{ohlsen}
   \end{eqnarray}
were $P_y$ ($P_y^T$) is the  vector polarization of the beam (target), $A_y$ ($A_Y^T$)  is the vector analyzing power in respect of the polarized beam (target) and $C_{y,y}$ is the spin-spin correlation coefficient; $I_0$ is the $I$ for unpolarized particles. We consider below two options for dd-collision, first one for $P_y=P_y^T=\frac{2}{3}$ and the second for  $P_y=\frac{2}{3}$, $P_y^T=-\frac{2}{3}$.
The cross section is denoted as $I_{\uparrow\uparrow}$ for the first option and as $I_{\uparrow\downarrow}$ for the second one. Using Eq.(\ref{ohlsen}) one can  easy
 find the  following relation
 \begin{eqnarray}
  C_{y,y}=\frac{(I_{\uparrow\uparrow}-I_{\uparrow\downarrow})+
  (I_{\downarrow\downarrow}-I_{\downarrow\uparrow}) }
  { (I_{\uparrow\uparrow}+I_{\uparrow\downarrow})+
  (I_{\downarrow\downarrow}+I_{\downarrow\uparrow})},
  \label{cyy}
 \end{eqnarray}
 where $I_{\uparrow\uparrow}$ and  $I_{\uparrow\downarrow}$  are defined by
 Eqs. (\ref {Nupup}), $I_{\uparrow\downarrow}=I_{\downarrow\uparrow}$ and
 $\cal N_{\downarrow\downarrow}$ is
\begin{eqnarray}
 {\cal N}_{\downarrow\downarrow}= L(2\cdot 2d\sigma_{-1,-1}+2d\sigma_{-1,0}+ 2d\sigma_{0,-1}+d\sigma_{0,0}).
\end{eqnarray}
Using these relations and taking into account  $d\sigma_{-1,0}=d\sigma_{0,+1}$
in  Eq. (\ref{dsigmll}), one can find from Eq. (\ref{cyy})
\begin{eqnarray}
 C_{y,y}=\frac{4}{9}A_{YY}^{NN}.
\end{eqnarray}
Therefore, $C_{y,y}$ coincides with $A_{YY}^{dd}$ defined by
Eq.(\ref{Ayy}).

 \section{Summary}

  The nucleon-nucleon elastic scattering is a basic process in  physics of hadrons and   atomic
nuclei.
Our understanding of the available  data on the
$pp$- elastic  scattering with both polarized
 protons at energies
$\sqrt{s_{NN}}= 3 - 6$ GeV at large  scattering angles $\theta_{cm}\sim 90^\circ$ is still not  enough clear.
New important  information on  the dynamics of NN-elastic scattering  can be obtained from the double polarized $pn$- elastic scattering. We have shown here that a measurement of the double spin-correlation in the
reaction $dd\to pnpn$ in the kinematic providing dominance of the double pole diagram depicted in
Fig. \ref{fig1} is  simultaneously a measurement  of the double spin correlation in $pN$ elastic scattering.
This relation  is true for any orientation of the  quantization axis  for the case  when the  S-wave dominates in the deuteron wave function. This dominance takes the place for the case when the final neutron $n_1(0)$ and the final proton $p_2(0)$
are flying
in the direction of the corresponding initial beam and have one half
of the  momentum of the corresponding initial deuteron.
These conditions
 can be used at SPD NICA experiment \cite{Abramov:2021vtu,Uzikov:2022fux}.
The contribution of the D-wave and  effects of the initial and final state interactions will be studied separately.

 Concerning the counting rate $N$ of this process one should note that differential cross section
 of the $pp$-elastic scattering at $\sqrt{s_{NN}}=5$ GeV and
 $\theta_{cm}=90^\circ$  is $\sim 10^{-2} \mu b/sr$
 \cite{Akerlof:1967zz}. For the luminosity
 $\sim 10^{29} cm^{-2}sec^{-1}$ in $pp$-collision \cite{Abramov:2021vtu} this corresponds to
 $N\sim 10^{-3}/sec$. However, for the scattering angle $\theta_{cm}=50^\circ$ this number increases by two orders of magnitude
 \cite{Larionov:2022gvn}.

{\bf Acknowledgements.} One of the coauthors (Yu.N.U.) is thankfull to  N.M. Piskunov for important comments on the properties
of polarized deuteron beams.    This work is supported in part by the grant of the Scientific Programme  of the Republic of Kazakhstan -- JINR  for the 2023 year.

\bibliographystyle{pepan}
\bibliography{pep-uzm}

\end{document}